\documentclass[aps,prd,twocolumn,groupedaddress,amsfonts,amssymb,
amsmath,showpacs,floatfix]{revtex4-1}
\usepackage{dcolumn}
\usepackage{multirow}
\usepackage{verbatim}
\usepackage{graphicx}
\usepackage{color}

\newcommand{\ms}{\overline{MS}}

\begin{document}

%\preprint{}

\title{Locally smeared operator product expansions in scalar 
field theory}


\author{Christopher \surname{Monahan}}
%\thanks{}
\affiliation{Physics Department, College of William and Mary,
Williamsburg, Virginia 23187, USA}
\altaffiliation{Department of Physics and Astronomy, University of Utah, Salt 
Lake City, Utah 84112, USA}
\author{Kostas \surname{Orginos}}
\affiliation{Physics Department, College of William and Mary,
Williamsburg, Virginia 23187, USA\\
and Thomas Jefferson National Accelerator Facility, Newport News, 
Virginia 23606, USA}
\author{}

\noaffiliation

%\date{\today}

\begin{abstract}
We propose a new locally smeared operator product expansion to 
decompose nonlocal operators in terms of a basis of smeared
operators. The smeared operator product expansion formally connects 
nonperturbative matrix elements 
determined numerically using lattice field 
theory to matrix elements of nonlocal operators in the continuum. These 
nonperturbative matrix 
elements do not suffer 
from power-divergent mixing on the lattice, which significantly 
complicates calculations of quantities such as the moments of parton 
distribution functions, provided the smearing scale is kept 
fixed in the continuum limit. The presence of this smearing scale complicates 
the connection to the Wilson coefficients of the standard operator product 
expansion and 
requires the construction of a suitable formalism. We demonstrate 
the feasibility of our approach with examples in real 
scalar field theory.
\end{abstract}

%\keywords{}

\maketitle

\section{Introduction}

Quantum chromodynamics (QCD) connects hadronic matter to its fundamental 
constituents, quarks, and gluons. Many aspects of QCD are poorly understood, in 
spite of four decades of intense 
experimental and theoretical effort. As part of this effort, deep inelastic 
scattering (DIS) of leptons from nucleons has played a central role in 
establishing QCD as the reigning theory of the strong interaction and continues 
to serve as a mainstay for attempts to unravel QCD's mysteries 
(for complete reviews see, for example, Refs.~\cite{pdg12} and 
\cite{Collins:2011zzd}).

Theoretically, inclusive DIS cross sections are 
separated into leptonic and hadronic tensors, which capture the 
electroweak and strong dynamics of the scattering process, respectively. The 
hadronic tensor factorizes into a convolution of infrared-safe perturbative 
coefficients and parton distribution functions (PDFs), 
which incorporate the low-energy QCD physics and therefore must be determined 
nonperturbatively. PDFs are independent of the scattering process--but 
depend on the target nucleon, while the perturbative coefficients are 
independent of the external scattering states.

PDFs are important for two reasons. First, they furnish 
direct knowledge of the constituents of hadronic states---the dominant form of 
visible matter in the universe---in terms of the fundamental theory of the 
strong force. Second, they provide constraints on hadronic 
backgrounds at collider experiments, such as the Large Hadron Collider. 
These backgrounds affect the sensitivity of a variety of high energy 
experiments, including studies of the properties of the Higgs boson 
\cite{Thorne:2011kq,Alekhin:2010dd,Alekhin:2011ey}, and searches for heavy $W'$ 
and $Z'$ boson production \cite{Brady:2011hb}.

PDFs cannot currently be calculated directly from QCD using \emph{ab 
initio} lattice QCD, because they are defined 
in terms of light-cone matrix elements that are not directly 
accessible on Euclidean lattices. Hence, PDFs
are usually extracted from global analyses of different experiments  
\cite{Jimenez-Delgado:2014xza,Jimenez-Delgado:2013boa,Ball:2013lla,
Arbabifar:2013tma,Owens:2012bv,Accardi:2011fa,Leader:2010rb,Blumlein:2010rn, 
Hirai:2008aj}. 

Naturally a direct nonperturbative computation of PDFs is desirable: with 
sufficient precision, such a calculation would further constrain global fits in 
regions that may be 
experimentally inaccessible.

Until recently, lattice calculations have focused on determining the Mellin 
moments of PDFs, which are related to matrix elements of twist-2
operators that \emph{can} be determined on the lattice 
\cite{Bali:2012av,Braun:2008ur,Guagnelli:2004ga}. The lattice regulator breaks 
Lorentz symmetry, which induces radiative mixing
between operators of different mass dimensions. The resulting power divergences 
in the lattice spacing completely 
obscure the continuum limit. Moments up to the fourth moment 
can 
be extracted by carefully choosing the external momenta and operators, but this 
significantly 
limits the precision with which one can 
extract meaningful results for PDFs \cite{Detmold:2003rq,Detmold:2001dv}. 
Beyond 
the fourth moment, power divergent mixing is inevitable and this method cannot 
provide reliable calculations.

Ji recently proposed a new approach to directly extract PDFs from lattice 
calculations \cite{Ma:2014jla,Ji:2013dva}, 
based on a large-momentum effective theory \cite{Ji:2014gla}. Preliminary 
results first appeared in Ref.~\cite{Lin:2014zya}, but a number of issues 
remain, including a complete understanding of the renormalization of the 
relevant lattice matrix elements and the practical ability to resolve 
sufficiently large momenta on the lattice.

Here we propose a new formalism that removes mixing in the continuum limit and, 
in 
principle, enables the extraction of higher moments of PDFs from lattice QCD. 
We call this formalism the ``smeared operator product expansion'' (sOPE). We 
expand continuum matrix elements in a basis of locally-smeared lattice 
degrees of 
freedom. The resulting matrix elements are functions of two scales, the 
smearing 
scale and the renormalization scale. The sOPE provides the framework necessary 
to relate these matrix elements, via Wilson coefficients, to 
phenomenologically useful quantities, such as the hadronic tensor of DIS.

Smearing has been widely used in lattice QCD to reduce ultraviolet 
fluctuations, partially restore rotational symmetry and thereby systematically 
improve the precision of lattice calculations 
\cite{Morningstar:2003gk,Hasenfratz:2001hp,Bernard:1999kc,Albanese:1987as}. For 
a pedagogical overview of smearing in lattice 
calculations, see, for example, Ref.~\cite{Gattringer:2010zz}. In the sOPE, we 
implement smearing 
via the gradient flow, a classical evolution of the fields in a new dimension, 
the flow time, toward the stationary points of the action 
\cite{Luscher:2013cpa,Luscher:2011bx,Lohmayer:2011si,Luscher:2010iy,
Narayanan:2006rf}. 
Matrix elements determined nonperturbatively on the lattice require no further 
renormalization, up to a fermionic wave function renormalization 
\cite{Luscher:2013cpa}, and remain 
finite, provided the smearing scale is kept fixed in physical units in the 
continuum limit \cite{Makino:2014sta,Luscher:2011bx}. There are two further 
advantages: the gradient flow allows one to use smearing lengths 
of only one or two lattice spacings, much smaller than hadronic length scales 
on typical lattices, and therefore does not distort the low energy physics 
\cite{Davoudi:2012ya}, and also the gradient flow is computationally very 
cheap. 

The gradient flow is now well-established as a tool to study lattice 
gauge theories, with applications from scale-setting, 
\emph{i.e.}~determining the lattice spacing in physical units 
\cite{Borsanyi:2012zr,Borsanyi:2012zs,Bazavov:2013gca,Fodor:2014cpa,
Cheng:2014jba}, to studying renormalization in 
lattice gauge theories. For example, 
the gradient flow has been used to define finite-volume renormalization schemes 
for the strong coupling constant
\cite{Fodor:2012td,Fritzsch:2013yxa,Fritzsch:2013hda, 
Ramos:2013gda,Rantaharju:2013bva}, for operator 
renormalization \cite{Monahan:2013lwa}, and to understand the nonperturbative 
scale dependence of renormalized matrix elements \cite{Luscher:2014kea}. 
Related work has used Ward identities to study chiral fermions on 
the lattice \cite{Shindler:2013bia} and the energy-momentum tensor 
\cite{Suzuki:2013gza,DelDebbio:2013zaa, Makino:2014taa}.

In this paper we introduce the gradient flow for a 
single real scalar field with quartic interactions and outline some of the 
properties of the sOPE applied to real scalar field theory.
Scalar field theories describe some important critical 
phenomena in nature, such as the antiferromagnetic phase transition, and 
have a long history as a testing ground for fundamental ideas in 
quantum field theory in four dimensions. For our purposes, the chief advantage 
is the simplicity of Euclidean scalar field theory, which lays bare the 
structure of the sOPE and highlights the points of similarity and contrast with 
the local operator product expansion (OPE). 

In the next section, we review Wilson's OPE and introduce the sOPE. We then 
apply the sOPE to scalar field theory in Sec.~\ref{sec:phi4}, illustrate how 
the gradient flow removes power-divergent mixing in Sec.~
\ref{sec:lattsmear}, 
and calculate 
the perturbative Wilson coefficients to two loops in Sec.~\ref{sec:wc}. 
Finally, in Sec.~\ref{sec:rgeqns}, we study the scale dependence of the sOPE 
through renormalization group 
equations for the Wilson coefficients.

\section{\label{sec:ope}The operator product expansion}

Wilson's approach to the OPE for a nonlocal operator is widely known---see, for 
example, Ref.~\cite{Collins:1984rdg}---and here we review some notation 
necessary for our discussion.

We write the OPE for a nonlocal operator, ${\cal Q}(x)$, as
\begin{equation}
{\cal Q}(x)  \stackrel{x\rightarrow 0}{\sim} \; \sum_k 
c_k(x,\mu) [{\cal O}^{(k)}(0)]_{\mathrm{R}}.
\end{equation}
The Wilson coefficients $c_k(x,\mu)$ are complex functions of the spacetime 
separation, $x$, and renormalization scale, $\mu$, that capture the 
short-distance physics associated with the renormalized local operator $[{\cal 
O}^{(k)}(0)]_{\mathrm{R}}$. We represent renormalized operators by 
$[\ldots]_{\mathrm{R}}$ and suppress their dependence on the renormalization 
scale, $\mu$, for notational simplicity. The 
free-field mass dimension of the local operator determines the leading 
spacetime dependence of the corresponding Wilson coefficient.

As an example, let us consider the time-ordered two-point function of two 
scalar fields with spacetime separation $x$: ${\cal T} \{\phi(x)\phi(0)\}$. In 
free scalar field theory, 
the OPE is a Laurent expansion. Interactions generate subleading dependence 
on the spacetime separation in the Wilson coefficients, which become 
functions of the spacetime separation, the (renormalized) mass 
$m_{\mathrm{R}}$, and the 
renormalization scale, $\mu$:
\begin{align}\label{eq:opeint}
{\cal T} \{\phi(x)\phi(0)\} = {}& \frac{c_{\mathbb{I}}(\mu 
x,m_{\mathrm{R}}x)}{4\pi^2x^2}\mathbb{I}+ 
c_{\phi^2}(\mu x,m_{\mathrm{R}}x)[\phi^2(0) ]_{\mathrm{R}} \nonumber\\
{} & \qquad 
+{\cal O}(x).
\end{align}
Here we have 
factored 
out 
the leading spacetime dependence from the Wilson 
coefficients. The ${\cal O}(x)$ indicates terms of order $x$, up to 
logarithmic corrections generated by interactions.

We propose replacing the set of local operators in the OPE by
their locally smeared counterparts
\begin{equation}
{\cal Q}(x)  \stackrel{x\rightarrow 0}{\sim} \; \sum_k 
d_k(\tau,x,\mu) {\cal S}^{(k)}(\tau,0).
\end{equation}
The Wilson coefficients, $d_k(\tau,x,\mu)$, and the smeared 
operators, ${\cal S}^{(k)}(\tau,0)$, are now functions of an extra 
scale, the 
flow time, $\tau$. Nevertheless, the leading spacetime 
dependence of the Wilson coefficients is dictated by the canonical mass 
dimension of the corresponding smeared operator
and is therefore just that of the standard OPE. 

Just as the OPE is only valid 
for small spacetime separations, the sOPE requires small flow times. We will 
see that we require that $\tau \propto x^2$, where $x$ is assumed to be small, 
to ensure that the Wilson coefficients are independent of the external states.

For example, returning to the time-ordered two-point 
function, the sOPE is
\begin{align}\label{eq:sope2pt}
\!\!{\cal T}\{\phi(x)\phi(0)\} ={} &
\frac{d_{\mathbb{I}}(\mu x,\mu^2\tau,m_{\mathrm{R}}x)}{4\pi^2x^2}\mathbb{I}
\nonumber\\
{} & + 
d_{\rho^2}(\mu x,\mu^2\tau,m_{\mathrm{R}}x)\rho^2(\tau,0)  + 
{\cal O}(x,\tau),
\end{align}
where we denote smeared fields at flow time $\tau$ by $\rho(\tau,x)$.

In the following sections, we will observe four features of the 
smeared Wilson coefficients in the sOPE.
\begin{enumerate}
\item The logarithmic spacetime dependence of the original OPE is preserved in 
the sOPE.
\item The flow time serves as an ultraviolet regulator for the smeared Wilson 
coefficients, to leading-order in perturbation theory. 
\item Beyond leading-order the flow time cannot regularize the Wilson 
coefficients, because the flow evolution is classical. We can, however, absorb 
renormalization scale dependence into the renormalization parameters 
of the original theory.
\item For any OPE to be meaningful, the Wilson coefficients must be independent 
of the external states. We can ensure this for the sOPE by choosing 
$s_{\mathrm{rms}} < x$; \emph{i.e.}~the mean smearing radius must be smaller 
than the spacetime extent of the nonlocal operator. This ensures that the sOPE
remains an expansion in approximately local operators. In other words, if the 
gradient flow probes length scales on the order of the nonlocal operator, then 
the sOPE becomes a poor expansion for the original operator.
\end{enumerate}

Although we refer to this expansion as the \emph{smeared} OPE, we really have 
in mind that the smearing is implemented via the gradient flow. The 
gradient flow acts as a smoothing operation that
drives the degrees of freedom of the theory to the stationary points of the 
action. In
QCD, the gradient flow corresponds to a continuous stout-smearing procedure, an 
analytic method for constructing lattice gauge fields with damped ultraviolet
fluctuations \cite{Morningstar:2003gk}. The direct comparison of the gradient 
flow to other smearing schemes was first undertaken in 
Ref.~\cite{Bonati:2014tqa}.

Many studies of the gradient flow have incorporated a small flow-time expansion 
of fields at nonvanishing flow time in terms of local 
fields at zero 
flow time \cite{Shindler:2013bia,Suzuki:2013gza,DelDebbio:2013zaa, 
Makino:2014taa,Asakawa:2013laa,Luscher:2014kea}. We can view such an expansion 
as an OPE in the flow time and thereby relate 
renormalized quantities calculated at nonzero flow time to the corresponding 
quantities in the original theory at vanishing flow time, which would otherwise 
be difficult to compute.

In this work, we take a different approach. We do 
not expand the flowed fields in terms of original fields at nonzero flow time, 
but rather take 
as the fundamental objects of study the (matrix elements of)
fields at positive flow time. Both approaches reflect the physically-motivated 
expectation that smearing 
scales much smaller than the physical scales of the system should not distort 
the physics in question. The small flow-time expansion quantifies any 
deviations 
from this expectation and, 
furthermore, decouples analytic calculations of Wilson coefficients in the 
continuum from lattice calculations with smeared degrees of freedom. The sOPE 
incorporates the smearing scale as an inherent scale of the system, which 
requires new Wilson coefficients to be determined. Thus, both the small 
flow-time 
expansion and the sOPE are shaped by the role of the smearing scale as 
ultraviolet regulator and are related, but distinct, conceptual approaches to 
the same physics: partially restoring rotational symmetry on the lattice.

Smearing, in 
general, is a tool that partially restores rotational symmetry 
\cite{Davoudi:2012ya} and thereby improves the continuum limit of lattice 
calculations. Smearing via the gradient flow has a number of advantages. In 
particular, 
matrix elements determined nonperturbatively on the lattice using 
smeared degrees of freedom require no further renormalization 
\cite{Luscher:2010iy}, up to fermionic 
wave function renormalization \cite{Luscher:2013cpa}, and remain 
finite, provided the smearing scale is kept fixed in physical units as the 
continuum limit is taken.

We now turn to a more complete study of the sOPE applied to $\phi^4$ scalar 
field theory, a particularly straightforward theory in which to examine 
the sOPE. We can solve the flow-time 
equations exactly, because the flow-time 
evolution is linear in the scalar field.
We do not need to consider the complications associated with 
gauge fixing \cite{Luscher:2010iy,Luscher:2011bx}---nor the extra 
renormalization of fermions \cite{Luscher:2013cpa}. Furthermore, for scalar 
fields the sOPE can be understood to be simply a resummation of the original 
OPE. Although it is not necessary for our work, it is also interesting to note 
that 
the OPE is known to converge for Euclidean $\phi^4$ theory in four dimensions, 
at a fixed, but arbitrary order in the perturbative 
expansion \cite{Hollands:2011gf}. This result holds at arbitrary spacetime 
separations, provided the external states are of compact support.

Looking toward future calculations in QCD, we anticipate that 
the technical issues associated with a flow equation that incorporates gauge 
field interactions will slightly complicate the perturbative calculations by 
increasing the number of diagrams at a given order in perturbation theory, but 
will not alter our conclusions. Within the gauge sector, there are no loops of 
flowed fields, because renormalized correlation functions remain finite at 
nonzero flow time \cite{Luscher:2011bx}. Therefore, at leading-order in 
perturbation theory, the flow time regulates ultraviolet divergences; beyond 
leading-order an appropriate renormalization procedure must be incorporated. 
With fermionic fields there is an extra fermion renormalization parameter, 
calculated in Ref.~\cite{Luscher:2013cpa}, but this can be removed by 
considering 
renormalization group invariant quantities \cite{Luscher:2013cpa}. We also note 
that all perturbative calculations for the sOPE can be carried out in the 
continuum and do not require lattice perturbation theory, which is generally 
more involved \cite{Monahan:2013dla,Capitani:2002mp}.

\section{\label{sec:phi4}Gradient flow for scalar field theory}

We work in four-dimensional Euclidean scalar field theory with quartic 
self-interactions and bare mass $m$, defined by the action
\begin{equation}
S_\phi[\phi] = \frac{1}{2}\int \mathrm{d}^4x\, \left[(\partial_\nu  
\phi)^2 + m^2\phi^2 + \frac{\lambda}{12}\phi^4\right].
\end{equation}

To study the sOPE, we introduce the scalar 
gradient flow, which we define through the flow-time evolution equation
\begin{equation}
\frac{\partial \rho(\tau,x)}{\partial \tau}= \partial^2 \rho,
\end{equation}
where $\rho$ is a scalar field at nonzero flow time, $\tau$, and 
$\partial^2$ is the Euclidean, four-dimensional Laplacian operator. 
Note that the flow time has units $[\tau] = [x]^2$.

Imposing the Dirichlet 
boundary condition 
$\rho(0,x) = \phi(x)$, we may 
write the exact solution of the flow time equation as
\begin{equation}\label{eq:rhoexp}
\rho(\tau,x) = e^{\tau\partial^2}\phi(x),
\end{equation}
or, in the momentum representation, as
\begin{equation}
\tilde{\rho}(\tau,p) = e^{-\tau p^2}\tilde{\phi}(p).
\end{equation}

The full solution is
\begin{align}
\rho(\tau,x) = {} & \int\mathrm{d}^4y\,\int \frac{\mathrm{d}^4p}{(2\pi)^4} 
\,e^{ip\cdot(x-y)} e^{-\tau p^2} \phi(y) \nonumber\\
= {} &  \frac{1}{16\pi^2\tau^2}
\int\mathrm{d}^4y\,e^{-(x-y)^2/4\tau}\phi(y),
\end{align}
which demonstrates explicitly the ``smearing'' effect of the gradient flow: the 
flow time exponentially suppresses ultraviolet modes.  We 
parametrize the smearing radius via the root-mean-square smearing length, 
$s_{\mathrm{rms}} = \sqrt{8\tau}$.

The (Euclidean) smeared scalar propagator, for two fields at flow times 
$\tau_1$ and $\tau_2$, is given by
\begin{equation}
\widetilde{G}_\rho(\tau_1,\tau_2,k) = \frac{e^{-(\tau_1+\tau_2) k^2}}{k^2+m^2}.
\end{equation}
The flow-time evolution is classical, so any interactions must occur at 
zero flow time. The corresponding Feynman rule for the four-point vertex is 
just that of the standard (Euclidean) four-point vertex, $V^{(4)} = 
-\lambda/24$.

\section{\label{sec:lattsmear} Mixing on the lattice}

Before we examine the sOPE in detail, we demonstrate how the gradient flow 
removes power-divergent mixing on the lattice. We consider the example of 
twist-2 
operators in scalar field theory, which are symmetric and traceless and given by
\begin{equation}
T_{\mu_1\ldots\mu_n}(x) = \phi(x)\partial_{\mu_1}\ldots 
\partial_{\mu_n}\phi(x) - \mathrm{traces}.
\end{equation}
On the lattice, we replace the partial derivatives with discrete difference 
operators
\begin{equation}
T_{\mu_1\ldots\mu_n}^{\mathrm{latt}}(x) = \phi(x)\Delta_{\mu_1}\ldots 
\Delta_{\mu_n}\phi(x) - 
\mathrm{traces}.
\end{equation}
The spacetime point $x$ is now a node in the lattice, $x_\mu = 
an_\mu$, where $a$ is the lattice spacing and $n_\mu$ is a four-component 
lattice vector. The discrete difference operators can be improved to remove 
discretization effects, but the simplest such symmetric operator acts on a 
scalar field as
\begin{equation}
\Delta_\mu \phi(x) = \frac{1}{2a}\left(\phi(x+\widehat{\mu}) - 
\phi(x-\widehat{\mu})\right),
\end{equation}
where $\hat{\mu}$ is the unit vector in the $\mu^{\mathrm{th}}$ direction.

The lattice regulator breaks rotational symmetry, which induces mixing between 
twist-2 operators of different mass dimension. On dimensional grounds, the 
mixing coefficients scale with inverse powers of the lattice spacing and these 
coefficients diverge in the continuum limit. This is the problem of 
power-divergent mixing on the lattice. For example, the simple
bilinear $T^{\mathrm{latt}}(x) = \phi(x) \phi(x)$ mixes with the 
operator $T_{\mu\nu}^{\mathrm{latt}}(x)$ with 
a coefficient that scales as $1/a^2$. More generally, the mixing between 
$T^{\mathrm{latt}}(x)$ and any twist-2 operator with an even number of 
derivatives, 
$T_{\mu_1\ldots\mu_{2n}}^{\mathrm{latt}}(x)$, scales as $1/a^{2n}$.

The smeared counterparts of these operators, which are given by
\begin{equation}
S_{\mu_1\ldots\mu_n}^{\mathrm{latt}}(x) = 
\rho(\tau,x)\Delta_{\mu_1}\ldots 
\Delta_{\mu_n}\rho(\tau,x) - 
\mathrm{traces}.
\end{equation}
do not suffer from this problem. The mixing coefficient between
$S^{\mathrm{latt}}(x) = \rho(\tau,x) \rho(\tau,x)$ and 
$S_{\mu_1\ldots\mu_{2n}}^{\mathrm{latt}}(x)$ scales as $1/\tau^{n}$, where 
$\tau$ is the flow time in 
\emph{physical units}. Provided we keep
the dimensionful scale $\tau=\widetilde{\tau}a^2$ fixed, where 
$\widetilde{\tau}$ 
is dimensionless, then as the lattice spacing $a$ decreases, the mixing 
coefficient remains finite, because matrix elements at nonzero flow time 
cannot contain any additional divergences \cite{Makino:2014sta,Luscher:2011bx}.

As a simple illustration of this behavior, let us consider the matrix element 
of two twist-2 operators of different mass dimension:
\begin{equation}
M_{\mathrm{cont}} = \left\langle \Omega | \phi(0) \phi(0) \,\phi(0) 
\partial_\mu\partial_\nu \phi(0) |\Omega \right\rangle.
\end{equation}
This matrix element vanishes for massless scalar fields. On 
the lattice, however, the corresponding matrix element,
\begin{equation}
M_{\mathrm{latt}} = \left\langle \Omega | \phi(0) \phi(0) \,\phi(0) 
\Delta_\mu\Delta_\nu \phi(0) |\Omega \right\rangle, 
\end{equation}
is nonzero.

\begin{figure}
\begin{minipage}{0.24\textwidth}
\includegraphics[width=0.9\textwidth,keepaspectratio=true]{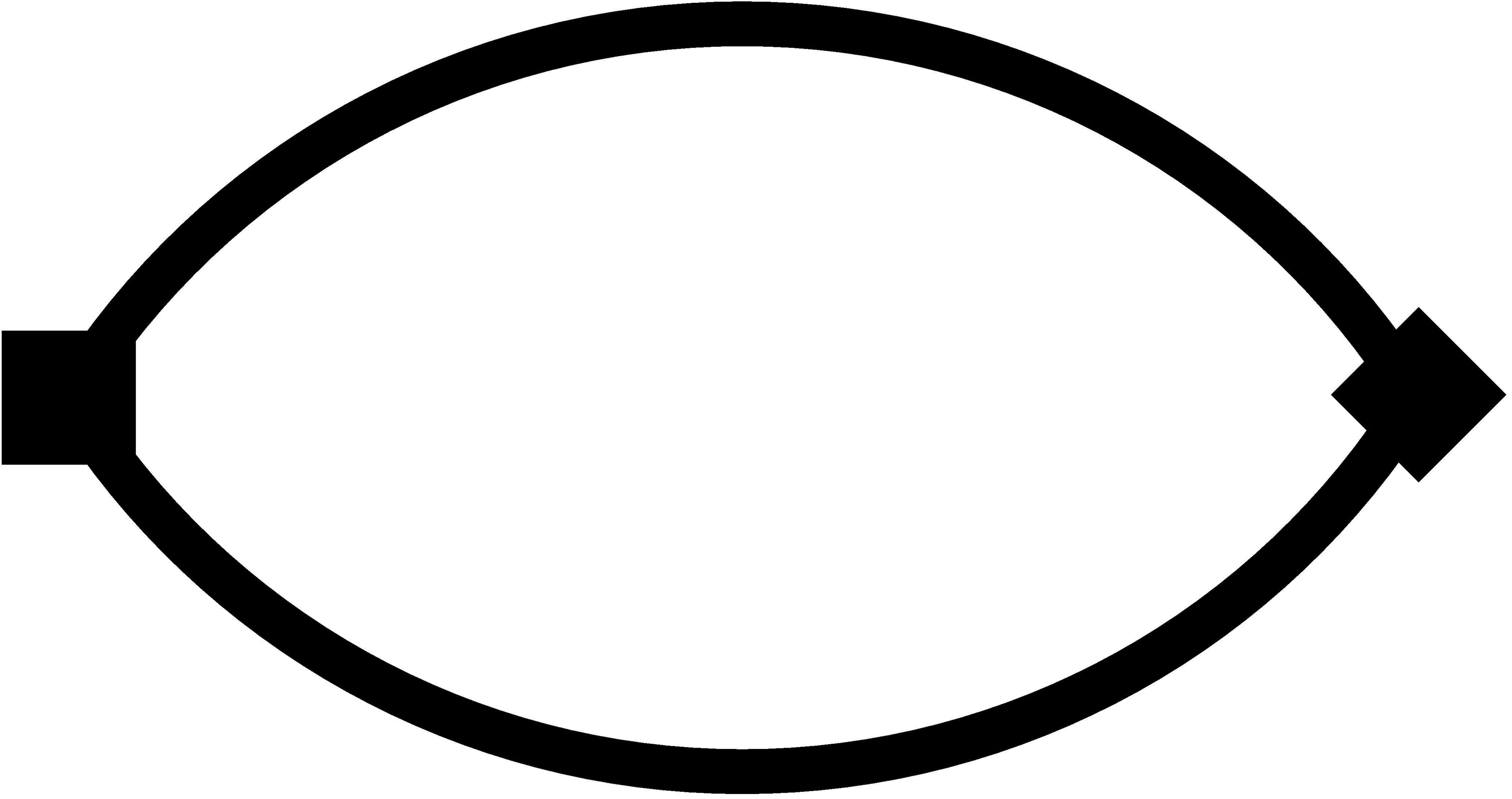}
\end{minipage}%
\begin{minipage}{0.24\textwidth}
\includegraphics[width=0.9\textwidth,keepaspectratio=true]{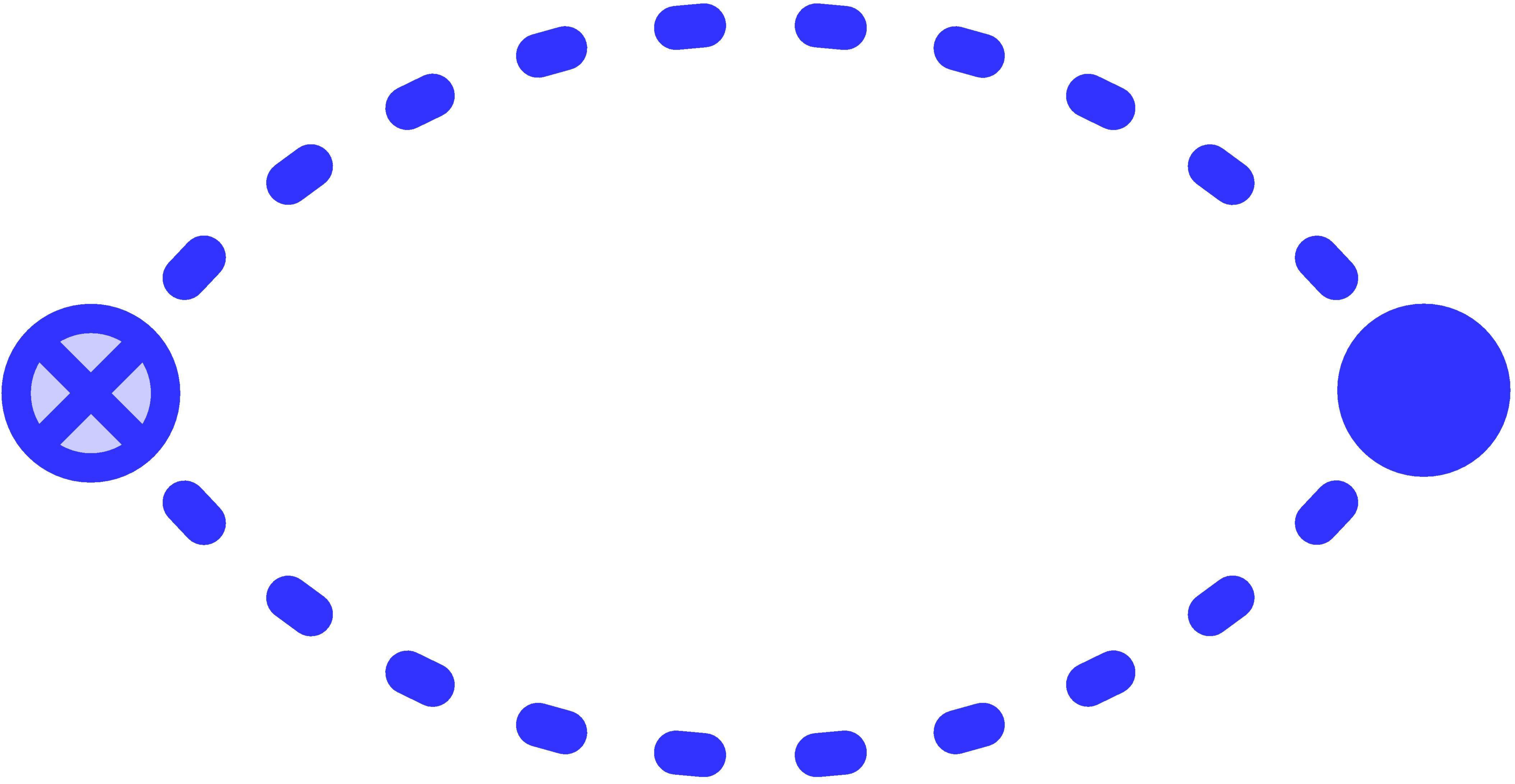}
\end{minipage}
\caption{\label{fig:mmix}Diagrams representing the leading contributions to the 
mixing matrix elements with unsmeared fields, $M_{\mathrm{latt}}^{(0)}$, and 
smeared fields, $M_{\mathrm{latt}}^{(0)}(\tau)$. The black solid lines in the 
left-hand diagram represent unsmeared propagators and the black solid square 
and 
diamond are the operators $\phi^2(0)$ and $\phi(0)\Delta_\mu\Delta_\nu\phi(0)$. 
The blue dashed lines in the right-hand diagram are smeared propagators and the 
two different blue blobs, filled and unfilled, represent the smeared operators 
$\rho^2(\tau,0)$ and $\rho(\tau,0)\Delta_\mu\Delta_\nu\rho(\tau,0)$.}
\end{figure}

We show the Feynman diagram for the leading contribution to this matrix element 
in the left-hand diagram of Fig.~\ref{fig:mmix}, which is given by
\begin{equation}
M^{(0)}_{\mathrm{latt}} = 
\frac{1}{a^4}\int^{\pi/a}_{-\pi/a}\frac{\mathrm{d}^4(ak)}{(2\pi)^4}\
\frac{\widehat{k}_\mu\widehat{k}_\nu}{\big(\widehat{k}^2+m_0^2\big)^2 } , 
\end{equation}
where $\widehat{k}_\mu = (2/a)\sin(ak_\mu/2)$. We have included a bare mass 
$m_0$ to remove spurious infrared divergences, but the integral is finite and 
we can take the massless limit, $m_0 \rightarrow 0$, to match to the continuum 
massless theory.
We expand this in powers of the lattice spacing to obtain
\begin{align}
M^{(0)}_{\mathrm{latt}} = {} & -\frac{\delta_{\mu\nu}}{4}
\int^{\pi/a}_{-\pi/a}\frac{\mathrm{d}^4k}{(2\pi)^4}
\frac{k^2}{(k^2+m_0^2)^2 } +{\cal O}(a^2) \nonumber\\
= {} & 
-\frac{\delta_{\mu\nu}}{48a^2}+{\cal 
O}(a^0)
\end{align}
in the massless limit. This result signals the appearance of power-divergent 
mixing: the $1/a^2$ term diverges in the continuum limit. Although this 
calculation is only leading-order in perturbation theory, higher-order 
contributions do not modify this power-divergent dependence on the lattice 
spacing \cite{Davoudi:2012ya}.

If we calculate this matrix element with smeared degrees of freedom 
(which we depict in the right-hand 
diagram of Fig.~\ref{fig:mmix}), however, we find
\begin{align}
M^{(0)}_{\mathrm{latt}}(\tau) = {} &\left\langle \Omega | \rho(\tau,0) 
\rho(\tau,0) \,\rho(\tau,0) 
\Delta_\mu\Delta_\nu \rho(\tau,0) |\Omega \right\rangle \nonumber\\ 
= {} &  -\frac{\delta_{\mu\nu}}{4}
\int^{\pi/a}_{-\pi/a}\frac{\mathrm{d}^4k}{(2\pi)^4}
\frac{k^2e^{-4\tau k^2}}{(k^2+m_0^2)^2 } +{\cal O}(a^2) \nonumber\\
= {} & \delta_{\mu\nu}
\frac{e^{-4\pi^2\tau/a^2}-1}{256\pi^2\tau}+{\cal 
O}(a^0).
\end{align}
Once again we have taken the massless limit. In the continuum limit, 
keeping the flow time $\tau$ fixed in physical units, this matrix element 
tends to a constant, signaling the suppression of power-divergent mixing for 
smeared degrees of freedom.

\section{\label{sec:wc}Wilson coefficients for the sOPE}

The procedure for calculating smeared Wilson coefficients parallels that 
for the OPE, discussed in, for example, Ref.~\cite{Collins:1984rdg}. With the 
Feynman rules of 
Sec.~\ref{sec:phi4} in hand, the calculation of smeared Wilson 
coefficients up to next-to-leading-order in perturbation theory is 
straightforward. We 
determine the smeared Wilson coefficients for the leading connected and 
disconnected operators, starting with the disconnected contribution.

\subsection{Disconnected contributions}

\subsubsection{Leading order}

\begin{figure}
\begin{minipage}{0.24\textwidth}
\includegraphics[width=0.9\textwidth,keepaspectratio=true]{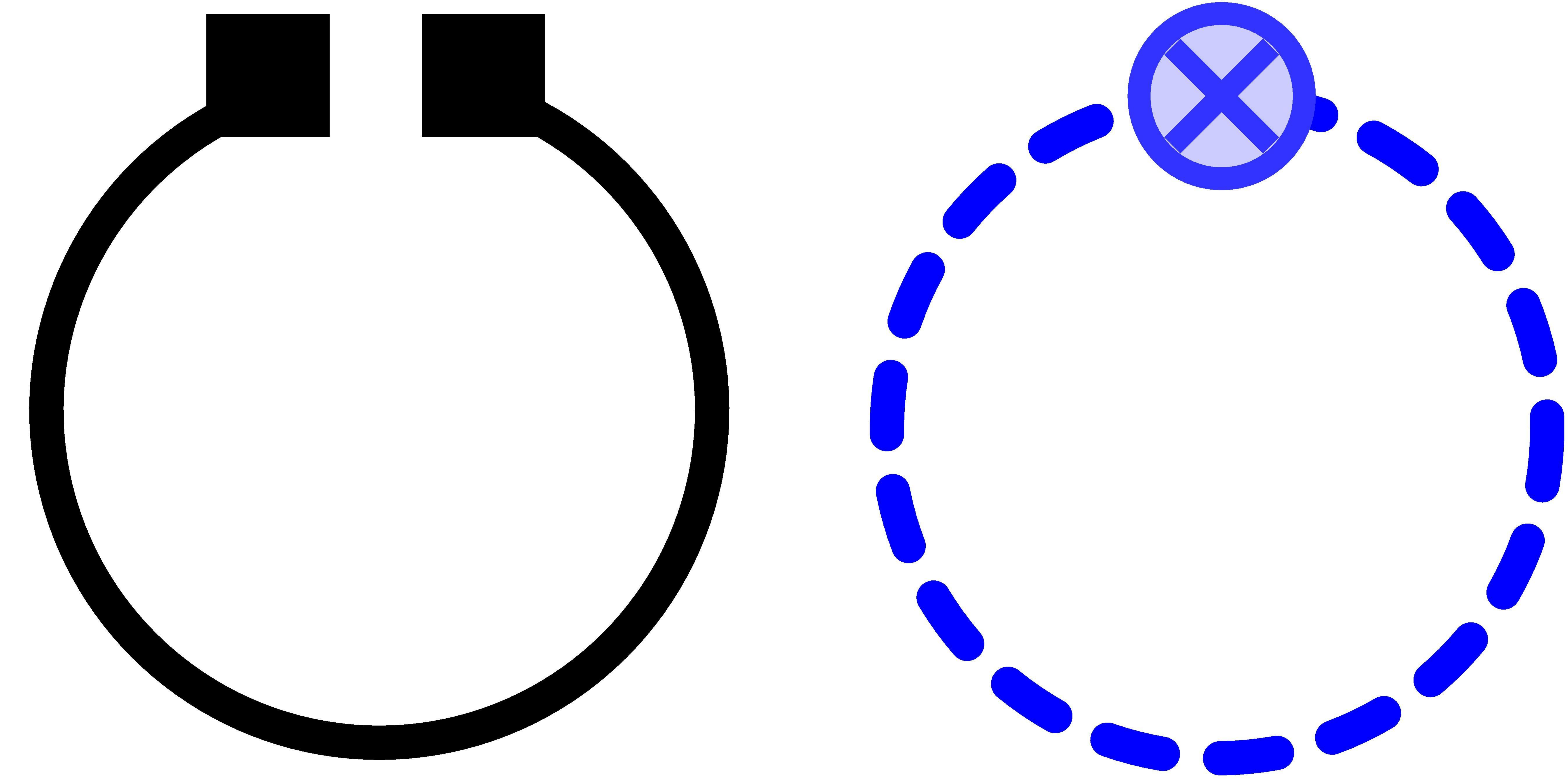}
\end{minipage}%
\begin{minipage}{0.24\textwidth}
\includegraphics[width=0.9\textwidth,keepaspectratio=true]{d1loop}
\end{minipage}
\caption{\label{fig:sope2pt}Diagrams representing the contributions to the 
Wilson coefficient $d_{\mathbb{I}}$ at leading-order (left two 
diagrams) and next-to-leading-order (right two diagrams). Solid black and 
dashed blue 
lines are propagators at vanishing and nonvanishing flow times, respectively. 
The black squares are unsmeared fields $\phi(0)$, black dots are 
interaction vertices at vanishing flow time and the blue blob represents 
the smeared operator $\rho^2(\tau,0)$.}
\end{figure}

We illustrate the leading-order (tree-level) and next-to-leading-order 
(one-loop) contributions to the smeared Wilson 
coefficient for the disconnected operator, $\mathbb{I}$, in Fig.~
\ref{fig:sope2pt}.
As we will see, the leading-order 
contribution is independent of the renormalization scale, $\mu$, because at 
this order the flow time serves as the ultraviolet regulator. Beyond leading 
order, however, the smeared Wilson coefficient will have a renormalization 
scale dependence.

We extract the disconnected smeared Wilson coefficient by considering matrix 
elements of each of the operators in the sOPE in the vacuum, which removes any 
connected contributions.  To ${\cal O}(x)$ we have
\begin{align}
\big\langle \Omega  |  {} &{\cal T}\{ 
\phi(x) \phi(0)\}  | \Omega \big\rangle =  
\frac{d_{\mathbb{I}}(\mu x,\mu^2\tau,mx)}{4\pi^2x^2}\big\langle \Omega | \,
\mathbb{I} \,| \Omega \big\rangle
\nonumber\\
{} & + 
d_{\rho^2}(\mu x,\mu^2\tau,mx)\big\langle \Omega | [\rho^2(\tau,0) 
]_{\mathrm{R}}| \Omega 
\big\rangle  + 
{\cal O}(x),
\end{align}
Here we have chosen the normalization of $d_{\mathbb{I}}$ so that at leading 
order in the free theory $d_{\mathbb{I}}$ is unity. We expand each quantity in 
this expression to one loop according to
\begin{equation}\label{eq:drhoexp}
f = f^{(0)} - \lambda f^{(1)} + {\cal O}(\lambda^2),
\end{equation}
where $f$ stands for either a matrix element or Wilson coefficient. We have 
chosen the sign of the one-loop contribution to factor out the sign of the 
coupling constant arising from the Feynman rule for the four-point vertex 
$V^{(4)} = -\lambda/24$. 

The tree-level Wilson coefficient is then given by the small spacetime 
behavior of
\begin{align}
\frac{d_{\mathbb{I}}^{(0)}}{4\pi^2x^2}  \stackrel{x\sim 0}{=}{} & 
 \Big\{\big\langle 
\Omega | 
{\cal T}\{ 
\phi(x){} \phi(0)\} | \Omega \big\rangle \nonumber\\
{} & \qquad - \big\langle \Omega | 
\rho^2(\tau,0)  | \Omega 
\big\rangle\Big\}_{{\cal O}(m^2)}^{(0)} ,
\end{align}
where we have neglected the arguments of the coefficient for clarity and used 
the fact that the tree-level connected
coefficient is $d_{\rho^2}^{(0)}=1$. The subscript indicates 
that we must expand the result to ${\cal O}(m^2)$. For more details about the 
calculation of Wilson coefficients see, for example, 
Ref.~\cite{Collins:1984rdg}. 
The corresponding Feynman integral representation is
\begin{align}
d_{\mathbb{I}}^{(0)} \stackrel{x\sim 0}{=} {} & 4\pi^2 x^2\left\{\int_k 
\frac{e^{ik 
\cdot 
x}-e^{-2k^2 \tau 
}}{k^2+m^2}\right\}_{{\cal O}(m^2)}
\nonumber \\
= {} & 4\pi^2x^2\int_k \left(e^{ik \cdot x} - e^{-2k^2 
\tau}\right)\left(\frac{1}{k^2} - \frac{m^2}{(k^2)^2}\right),
\end{align}
where
\begin{equation}
\int_k \equiv \int\frac{\mathrm{d}^4k}{(2\pi)^4}.
\end{equation}

The smeared Wilson 
coefficient is
\begin{equation}\label{eq:d1lead}
d_{\mathbb{I}}^{(0)} = 1 -
\frac{x^2}{8\tau}  +\frac{m^2x^2}{4}\left[\gamma_{\;\mathrm{E}} -1 + 
\log\left(\frac{x^2}{8\tau}\right)\right],
\end{equation}
with $\gamma_{\;\mathrm{E}}\simeq 0.577216$ the Euler--Mascheroni constant.

We can compare this result with the Wilson coefficient 
for the OPE for $d = 4-2\epsilon$ 
dimensions \cite{Collins:1984rdg}:
\begin{equation}
\overline{c}_{\mathbb{I}}^{(0)} =  
1 + \frac{  
m^2x^2}{4} \left[\frac{1}{\epsilon} + 1 +\gamma_{\;\mathrm{E}} + 
\log\left(\frac{\pi\mu^2x^2}{4}\right)\right].
\end{equation}
Then in the $\ms$ scheme, this becomes
\begin{equation}
\overline{c}_{\mathbb{I}}^{(0)} =  
1 +\frac{  
m^2x^2}{4} \left[1 +  2\gamma_{\;\mathrm{E}} + 
\log\left(\frac{\mu^2x^2}{16}\right)\right].
\end{equation}

Although the finite contribution to these expressions cannot be directly 
compared, 
because we have expressed the Wilson coefficients in two different 
renormalization schemes, we note three important features. First, the 
logarithmic dependence on the spacetime separation is identical. Second, 
the flow time $\tau$ plays the role of the renormalization scale at leading 
order. Third, we see that for small spacetime 
separations, we require a small flow-time parameter. If we do 
not choose the flow-time parameter appropriately, we generate large logarithmic 
contributions to the smeared Wilson coefficients and the sOPE exhibits poor 
convergence properties, even at small spacetime separations.

\subsubsection{Next-to-leading-order}

At one loop, the smeared Wilson coefficient is given by
\begin{align}
-\lambda\frac{d_{\mathbb{I}}^{(1)}}{4\pi^2x^2} = {} & \Big\{\big\langle 
\Omega | 
{\cal T}\{ 
\phi(x){} \phi(0)\} | \Omega \big\rangle \nonumber\\
{} & \qquad \qquad  - \big\langle \Omega | 
\rho^2(\tau,0)  | \Omega 
\big\rangle\Big\}_{{\cal O}(m^2)}^{(1)}.
\end{align}

The four-point interaction in this diagram, which we show in Fig.~
\ref{fig:sope2pt}, appears at zero flow time. Therefore the flow 
time cannot act as a regulator for the momentum integral over $k_2$ and we must 
introduce a renormalization procedure. We use dimensional regularization and 
the $\ms$ scheme. The double integral is straightforward, however, because the 
two integrals can be carried out separately:
\begin{equation}\label{eq:d11loop}
\frac{d_{\mathbb{I}}^{(1)}}{4\pi^2x^2}=  \bigg\{\int_{k_1}
\frac{e^{ik_1 \cdot 
x} - e^{- 2k_1^2\tau}}{(k_1^2+m^2)^2}  
\frac{1}{2}\int_{k_2}\frac{1}{k_2^2+m^2}\bigg\}_{{
 \cal O}(m^2)}.
\end{equation}
We find, for $m>0$,
\begin{align}
d_{\mathbb{I}}^{(1)} = {} & -
\frac{m^2x^2}{128\pi^2}\left[\gamma_{\;\mathrm{E}}-1+\log\left(\frac{x^2}{
8\tau } \right)\right]\left[
1+ \log\left(\frac{\mu^2}{m^2}\right)\right].
\end{align}

Here we see that the second term, which is a function of the new 
renormalization scale, $\mu$, and the bare mass, $m$, is nothing other than the 
one-loop contribution to the mass renormalization of the original theory, in 
the $\overline{MS}$ scheme \cite{Kleinert:2001ax}. Moreover, the factor 
containing the flow time 
is identical to the ${\cal O}(m^2)$ term from the leading-order contribution, 
$d_{\mathbb{I}}^{(0)}$, in Eq.~\eqref{eq:d1lead}.

Therefore, we can simply combine the leading-order and next-to-leading-order 
terms to give
\begin{align}
d_{\mathbb{I}} = {} & 1 -
\frac{x^2}{8\tau}  +\frac{m_{\mathrm{R}}^2x^2}{4}\left[\gamma_{\;\mathrm{E}} -1 
+ 
\log\left(\frac{x^2}{8\tau}\right)\right]  +{\cal O}(\lambda^2),
\end{align}
where $m_{\mathrm{R}}$ is the renormalized mass in the 
$\overline{MS}$ scheme, given by $m^2_{\mathrm{R}} = Z_m^{-1}m^2$ and $Z_m$ is 
the mass renormalization \cite{Kleinert:2001ax}:
\begin{equation}
Z_m = 1+\frac{\lambda}{16\pi^2}\left[
1+ \log\left(\frac{\mu^2}{m^2}\right)\right]+{\cal O}(\lambda^2).
\end{equation}

This is a clear, next-to-leading-order example of how the 
divergences of the theory at nonzero flow time are absorbed by 
the renormalization parameters of the original theory at zero flow time. In 
other words, the renormalized theory at zero flow time remains ultraviolet 
finite at nonzero flow time.

\subsection{Connected contribution}

\begin{figure}
\includegraphics[width=0.2\textwidth,keepaspectratio=true]{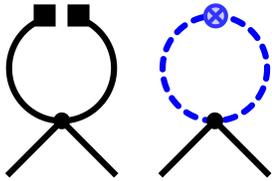}
\caption{\label{fig:sopeconn} 
Diagrams representing the leading-order (one-loop) contributions to 
the Wilson coefficient $d_{\rho^2}$. Details of the 
Feynman diagrams provided in the caption of Fig.~\ref{fig:sope2pt}.}
\end{figure}

\begin{figure}
\begin{minipage}{0.48\textwidth}
\includegraphics[width=0.9\textwidth,keepaspectratio=true]{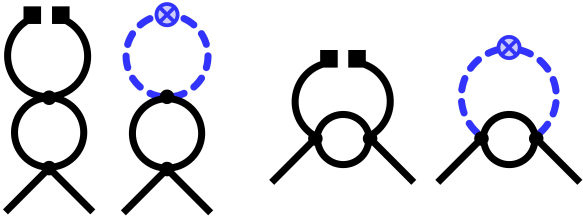}
\end{minipage}%
\caption{\label{fig:sopehigher} 
Diagrams representing the contributions to the next-to-leading-order 
(two-loop) contributions to the 
Wilson coefficients $d_{\rho^2}$. For details of the Feynman 
diagrams, see the caption of Fig.~\ref{fig:sope2pt}.}
\end{figure}

\begin{figure}
\includegraphics[width=0.48\textwidth,keepaspectratio=true]{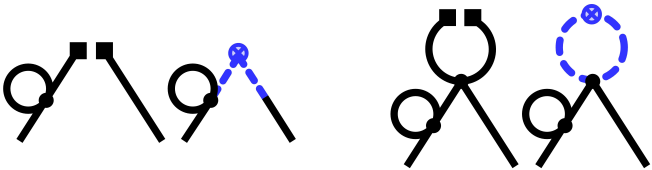}
\caption{\label{fig:zphi}
Example diagrams that are naturally 
incorporated in the renormalized external propagators. We do not include these 
contributions in our explicit calculations of the Wilson coefficients. For 
details of the Feynman 
diagrams, see the caption of Fig.~\ref{fig:sope2pt}.}
\end{figure}

We illustrate the leading- and next-to-leading-order contributions
to the connected Wilson coefficient $d_{\rho^2}(\mu x,\mu^2\tau,mx)$ of 
Eq.~\eqref{eq:sope2pt} in Figs.~\ref{fig:sopeconn} and 
\ref{fig:sopehigher}, respectively. 
Throughout this section, we neglect
diagrams that are trivially incorporated as part of the wave function 
renormalization of the external fields, such as the one-loop examples 
illustrated in Fig.~\ref{fig:zphi}. Provided the original 
theory at zero flow time is renormalized, counterterms cancel these 
contributions completely.

\subsubsection{Leading order}

We extract the leading-order connected Wilson coefficient by considering matrix 
elements of each of 
the operators in the sOPE coupled to two external fields, which removes any 
disconnected contributions.
We can then read off the one-loop contribution to the Wilson coefficient, 
$d_{\rho^2}^{(1)}$, by matching terms at ${\cal O}(\lambda)$. This 
contribution is given by the small spacetime behavior of
\begin{align}
-\lambda d_{\rho^2}^{(1)} {} & \stackrel{x\sim 0}{=}  
\Big\{ \big\langle \Omega | {\cal T}\{ 
\phi(x){} \phi(0)\}  
\widetilde{\phi}(p_1) \widetilde{\phi}(p_2) | \Omega \big\rangle^{(1)} 
\nonumber\\
{} & - \big\langle \Omega | \rho^2(\tau,0) 
\widetilde{\phi}(p_1) \widetilde{\phi}(p_2) | \Omega 
\big\rangle^{(1)}\Big\}_{{\cal O}(m^0)}.
\end{align}  

The corresponding Feynman integral 
is
\begin{align}\label{eq:curlyb}
\frac{1}{(p^2_1+m^2)(p_2^2+m^2)}\left\{\frac{1}{2}\int_k 
\frac{e^{ik\cdot x} - e^{-(k^2+q^2) \tau }}{(k^2+m^2)(q^2+m^2)}\right\},
\end{align}
where $q = k-p_1-p_2$. We extract the smeared Wilson 
coefficient by examining the small spacetime behavior of the 
integral in curly braces, expanded to ${\cal O}(m^0)$:
\begin{align}\label{eq:drho1}
d_{\rho^2}^{(1)} \stackrel{x\sim 0}{=} \left\{
\frac{1}{2}\int_k 
\frac{e^{ik\cdot x} - e^{-(k^2+q^2)\tau }}{(k^2+m^2)(q^2+m^2)}\right\}_{{\cal 
O}(m^0)}.
\end{align}

The 
smeared Wilson coefficients must be 
independent of the external states to ensure that the sOPE is truly an operator 
expansion. In this particular case, 
we require that 
$d_{\rho^2}^{(1)}$ is independent of the external momenta 
$p_1$ and $p_2$ and the mass. By taking 
a derivative with respect to one of the external momenta,
\begin{equation}\label{eq:drhoderiv}
\frac{\mathrm{d}}{\mathrm{d}p_i}d_{\rho^2}^{(1)} =
 \int_k \frac{q_i\big[e^{ik\cdot x} - 
e^{-(k^2+q^2)\tau}(1+(q^2+m^2)\tau)\big]}{(k^2+m^2)(q^2+m^2)^2},
\end{equation}
we obtain a convergent integral. The $x\rightarrow 0$ limit of this integral is 
now well defined and only vanishes if the flow time is 
related to the spacetime separation. An analogous result holds if we take a 
derivative with respect to the external mass, $m$. On dimensional grounds, 
then, we 
guarantee that the smeared Wilson coefficient is independent of the external 
states by choosing $\tau \propto x^2$.

\begin{figure}
\includegraphics[width=0.3\textwidth,keepaspectratio=true]{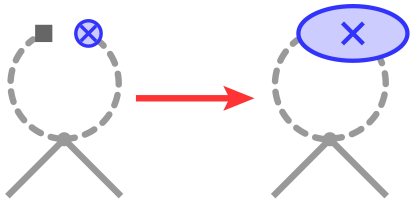}
\caption{\label{fig:sopeexpansion} At small flow times (left-hand diagram), 
the smeared operators are localized relative to the spacetime separation of 
the nonlocal operator. At large flow-time values (right-hand diagram), 
however, 
the smearing radius probes the scale of the nonlocal and the sOPE is a poor 
approximation to the original nonlocal operator. For details of the Feynman 
diagrams, see the caption of Fig.~\ref{fig:sope2pt}.
}
\end{figure}

On physical 
grounds, however, we require that the smearing radius is 
smaller than the spacetime extent of the nonlocal operator: 
$s_{\mathrm{rms}}< x$. This choice ensures that the sOPE remains an expansion 
in local operators. Physically speaking, if the 
gradient flow probes length scales on the order of 
the nonlocal operator, which we represent in Fig.~\ref{fig:sopeexpansion}, 
then smeared operators would cease to be (approximately) local. In other words, 
the 
sOPE would be a poor expansion for the original operator. This is the 
physical origin of the third feature that we saw in the leading-order 
disconnected contribution, Eq.~\eqref{eq:d1lead}: small spacetime 
separations require small flow times to ensure convergence. The OPE requires 
small spacetime 
separations for good convergence, 
so it follows that the sOPE requires small flow times as well. 

There is a complementary viewpoint that elucidates the small flow-time 
requirement more quantitatively: the small flow-time expansion 
\cite{Makino:2014taa,Suzuki:2013gza,Luscher:2011bx}. In this case we see that 
the 
derivative of Eq.~\eqref{eq:drhoderiv} is independent of the external 
momenta in the $x\rightarrow 0$ limit, up to terms linear in the flow time:
\begin{equation}
\frac{\mathrm{d}}{\mathrm{d}p_i}d_{\rho^2}^{(1)} =
\int_k\frac{q_i\big(e^{ik\cdot x} - 
1\big)}{(k^2+m^2)(q^2+m^2)^2}+{\cal O}(\tau).
\end{equation}
From this, it is clear that the Wilson coefficients will be independent of the 
external states, up to terms linear in the flow time. Therefore, provided the 
flow time is small relative to the spacetime separation, the sOPE is an 
expansion in approximately local operators, in a quantifiable sense. Moreover, 
the 
flow-time dependence will cancel in the product of the Wilson coefficient and 
its associated matrix element, to the desired order in the flow time.

We are free to set $p_1 = p_2 = 0$ in Eq.~
\eqref{eq:drho1}, because the smeared Wilson coefficient 
is independent of the 
external momenta and the mass to the order at which we are 
working \cite{Collins:1984rdg}. Expanding in the mass, we obtain
\begin{align}
d_{\rho^2}^{(1)} = {} &  
\frac{1}{2}\int_k 
\frac{e^{ik\cdot x} - e^{-2k^2 \tau }}{(k^2)^2} \nonumber \\
 = {} & -\frac{1}{32\pi^2}\left[\gamma_{\;\mathrm{E}} -1 + 
\log\left(\frac{x^2}{8\tau}\right)\right].
\end{align}
Combining this with the leading-order contribution, which is just unity, we have
\begin{equation}\label{eq:drhores}
d_{\rho^2}= 1 +\frac{\lambda}{32\pi^2}\left[\gamma_{\;\mathrm{E}} -1 + 
\log\left(\frac{x^2}{8\tau}\right)\right]+{\cal O}(\lambda^2).
\end{equation}

In contrast, the Wilson coefficient for the OPE in the $\ms$ scheme, denoted by 
$\overline{c}$, is
\begin{equation}\label{eq:cphires}
\overline{c}_{\phi^2} =  1+\frac{\lambda}{32\pi^2} \left[1 
+2\gamma_{\;\mathrm{E}}+ 
\log\left(\frac{\mu^2x^2}{16}\right)\right]+{\cal O}(\lambda^2).
\end{equation}
We note the occurrence of the three features we observed for the leading-order 
disconnected contribution: the same spacetime dependence for both smeared and 
unsmeared coefficients; the appearance of the flow time as a leading-order 
regulator; and the need to choose a small flow time for small spacetime 
separations to avoid large logarithmic contributions. From the small flow-time 
expansion viewpoint, we can confirm that the flow-time 
dependence ultimately cancels to the desired order. For example, the matrix 
element of $\rho^2(\tau,0)$ coupled to two external fields is
\begin{align}
\!\!\!\!\big\langle \Omega 
|{} & \rho^2(\tau,0)\widetilde{\phi}(p_1) \widetilde{\phi}(p_2)
|\Omega \big\rangle = 1  \nonumber\\
{} &\quad +\frac{\lambda}{32\pi^2}\left[1+\gamma_{\;\mathrm{E}} + 
\log\left(2m^2\tau\right)\right]
+{\cal O}(\tau,\lambda^2)
\end{align}
for sufficiently small flow times. Here we have dropped the external fields, 
which we take to be on shell, for simplicity. The product of this 
matrix element with the Wilson coefficient is independent of the flow time to 
one loop and ${\cal O}(\tau)$, as we would expect:
\begin{align}
d_{\rho^2}\big\langle  \Omega 
|{} &\rho^2(\tau,0)\widetilde{\phi}(p_1)\widetilde{\phi}(p_2)
| \Omega \big\rangle =  1 \nonumber\\
{} & 
+\frac{\lambda}{32\pi^2}\left[2\gamma_{\;\mathrm{E}} + 
\log\left(\frac{m^2x^2}{4}\right)\right]+{\cal O}(\tau,\lambda^2).
\end{align}

\subsubsection{Next-to-leading-order}

\begin{figure}
\begin{minipage}{0.3\textwidth}
\includegraphics[width=0.9\textwidth,keepaspectratio=true]{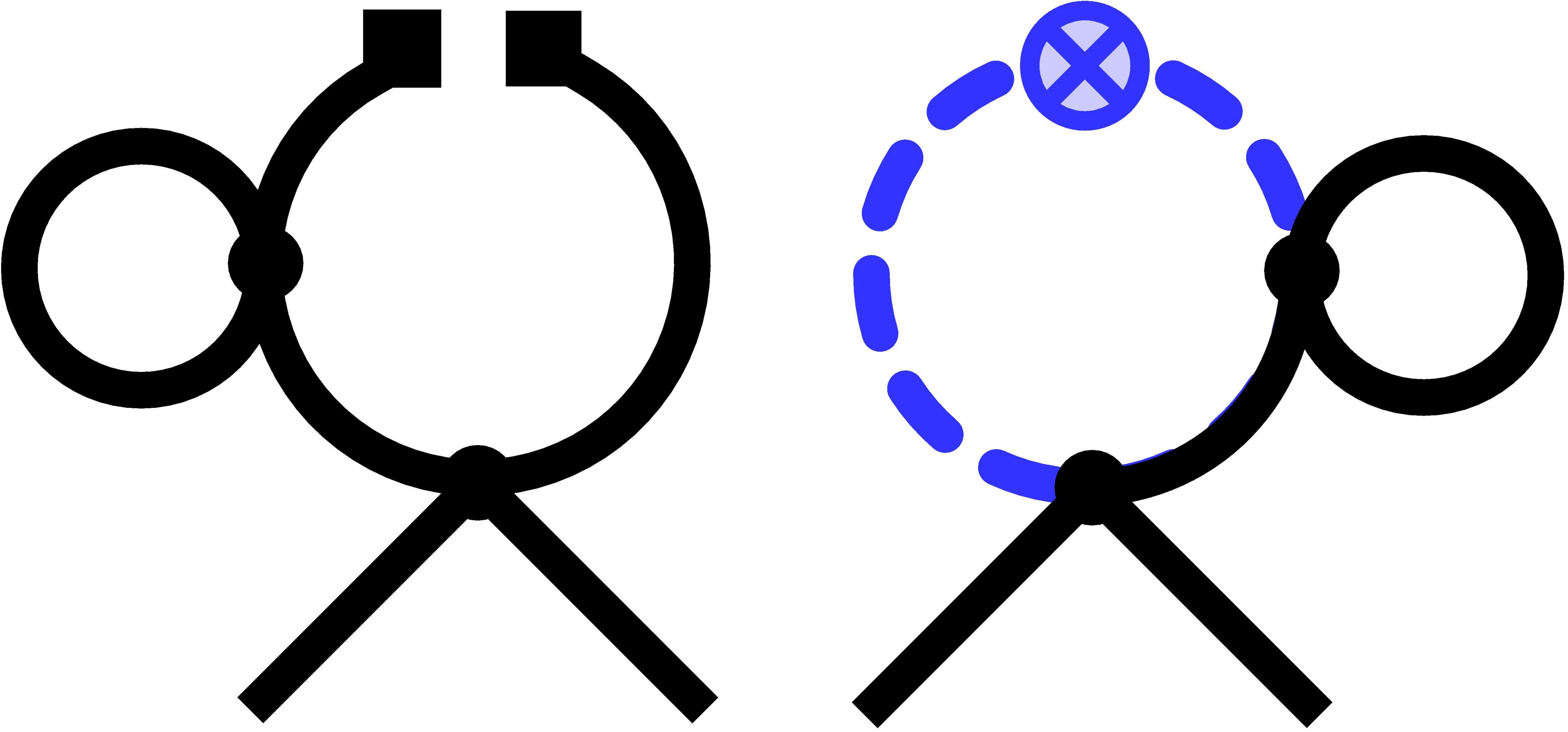}
\end{minipage}
\caption{\label{fig:2loopexc} 
Two-loop diagrams that appear at ${\cal O}(m^2)$ and therefore do not 
contribute to $d_{\rho^2}$. For details of the Feynman 
diagrams, see the caption of Fig.~\ref{fig:sope2pt}.}
\end{figure}

To determine the next-to-leading-order contribution to the connected Wilson 
coefficient, we must incorporate the Feynman diagrams of 
Fig.~\ref{fig:sopehigher}. The two-loop diagrams of Fig.~\ref{fig:2loopexc} do 
not 
contribute to the 
Wilson coefficient $d_{\rho^2}$, because they appear at ${\cal O}(m^2)$.

Looking at the Feynman diagrams of Fig.~\ref{fig:sopehigher}, and bearing 
in mind our experience of the next-to-leading-order contribution to 
$d_{\mathbb{I}}$, we immediately observe that these diagrams are simply the 
product of the one-loop Wilson coefficient, $d_{\rho^2}^{(1)}$, and the 
next-to-leading-order renormalized four-point vertex. This contribution to the 
vertex is, of course, nothing other than the next-to-leading-order contribution 
to the renormalized coupling constant. Thus, we write the Wilson 
coefficient quite simply as
\begin{equation}\label{eq:drho2r}
d_{\rho^2} = 1 +\frac{\lambda_{\mathrm{R}}}{32\pi^2}\left[\gamma_{\;\mathrm{E}} 
-1 + 
\log\left(\frac{x^2}{8\tau}\right)\right] +{\cal O}(\lambda^3),
\end{equation}
where $\lambda_{\mathrm{R}}$ is the renormalized coupling. In the $\ms$ scheme, 
the renormalized coupling is given by 
\begin{equation}\label{eq:lambdar}
\lambda_{\mathrm{R}} = \lambda - 
\frac{3\lambda^2}{2(16\pi^2)}\log\left(\frac{\mu^2}{m^2}\right)+{\cal 
O}(\lambda^3).
\end{equation}
For a calculation of this to five loops, see \cite{Kleinert:2001ax}.

As we move beyond the leading-order Wilson coefficients, 
\emph{i.e.}~$d_{\mathbb{I}}^{(0)}$ and $d_{\rho^2}^{(1)}$, divergent radiative 
corrections appear in our calculations, because all field interaction vertices 
appear at zero flow time. These divergences can be removed by the 
renormalization parameters of the original 
theory and the renormalization scale 
dependence of the smeared operators is completely contained in the 
renormalization parameters of the original theory. This is to be 
expected: it follows from the fact that renormalized matrix elements at zero 
flow time remain finite at nonzero flow time and require no further 
renormalization \cite{Makino:2014sta,Luscher:2011bx}.

Ultimately, for realistic calculations in lattice QCD, the 
perturbative 
calculation of smeared Wilson coefficients must be combined with 
nonperturbative determinations of matrix elements at hadronic energy 
scales. Scalar $\phi^4$ theory is not asymptotically free in four dimensions, 
but we can understand the mathematical features of the sOPE in more detail by 
studying the renormalization group equations for the simple example of 
scalar fields.

\section{\label{sec:rgeqns}Renormalization group equations}

We consider the matrix elements of scalar operators coupled to 
$N$ external, unsmeared scalar fields. We can derive renormalization group 
equations 
for the Wilson coefficients of the OPE by considering the scale dependence of 
suitably chosen Green functions \cite{Collins:1984rdg}. The Green function for 
$N+2$ external scalar 
fields obeys the renormalization group equation
\begin{equation}\label{eq:rg1}
\left[\mu\frac{\mathrm{d}}{\mathrm{d} \mu}+\left(N+2\right)\gamma 
\right]\big\langle \Omega |  
\widetilde{\phi}(p_1) \ldots\widetilde{\phi}(p_{N+2}) | \Omega \big\rangle = 0,
\end{equation}
while the Green function of the renormalized operator $\phi_{\mathrm{R}}^2(0)$ 
coupled 
to $N$ external scalar fields satisfies
\begin{equation}\label{eq:rg2}
\left[\mu\frac{\mathrm{d}}{\mathrm{d} \mu}-\gamma_m + N\gamma
\right]\big\langle \Omega |  \phi_{\mathrm{R}}^2(0)
\widetilde{\phi}(p_1) \ldots\widetilde{\phi}(p_N) | \Omega \big\rangle = 0.
\end{equation}
Here the renormalization group operator for scalar field theory is
\begin{equation}
\mu\frac{\mathrm{d}}{\mathrm{d} \mu} = \mu\frac{\partial}{\partial 
\mu}\bigg|_{\lambda,m_{\mathrm{R}}} + \beta 
\frac{\partial}{\partial \lambda}\bigg|_{\mu,m_{\mathrm{R}}} - \gamma_m 
m_{\mathrm{R}}\frac{\partial}{\partial m_{\mathrm{R}}}\bigg|_{\mu,\lambda}
\end{equation}
and the coefficients are \cite{Kleinert:2001ax}
\begin{align}
\beta^{\overline{MS}}(\lambda) = {} & \mu 
\frac{\mathrm{d}\lambda}{\mathrm{d}\mu} 
=\frac{3\lambda^2}{(16\pi^2)^2}+{\cal O}(\lambda^3), \\
\gamma_m^{\overline{MS}}(\lambda) = {} & 
-\frac{\mu}{2}\frac{\mathrm{d}\log(m_{\mathrm{R}})}{\mathrm{d}\mu}
\nonumber \\
= {} &  -\frac{\lambda}{2(16\pi^2)}+
\frac{5\lambda^2}{12(16\pi^2)^2}+{\cal O}(\lambda^3), \label{eq:gammam}\\
\gamma^{\overline{MS}}(\lambda) = {} & \frac{\mu}{2} 
\frac{\mathrm{d}\log(Z_\phi)}{\mathrm{d}\mu} 
=  \frac{\lambda^2}{12(16\pi^2)^2}+{\cal O}(\lambda^3).
\end{align}
In general these coefficients depend on the mass, the renormalization scale, 
and 
the renormalized coupling constant, but in a mass independent renormalization 
scheme, such as the $\overline{MS}$ scheme, they depend on the 
renormalization scale only through the renormalized coupling constant. These 
functions are known to five loops for the $O(N)$-symmetric theory, given by the 
$N$-multiplet $\Phi = \{\phi_1,\ldots,\phi_N\}$ 
\cite{Derkachov:1997pf,Kleinert:2001ax}.

Returning again to our example of the OPE for the two-point function, Eq.~
\eqref{eq:opeint}, we can derive an renormalization group equation for the 
Wilson coefficient 
$c_{\phi^2}$ by coupling these operators to two external scalar fields and 
using Eqs.~\eqref{eq:rg1} and \eqref{eq:rg2} (for further details, see, for 
example, Ref.~\cite{Collins:1984rdg}):
\begin{equation}
\left[\mu\frac{\mathrm{d}}{\mathrm{d} \mu} 
 + 2\big(\gamma +\gamma_m\big)\right]c_{\phi^2} = 0.
\end{equation}

Just as we might expect, the anomalous dimension of the Wilson coefficient 
$c_{\phi^2}$ is equal to the difference between the anomalous dimension of 
$\phi(x)\phi(0)$ and that of $[\phi^2(0)]_{\mathrm{R}}$.

\subsection{Renormalization group equations for the sOPE Wilson coefficients}

The flow time, $\tau$, introduces a new scale into the problem. In principle 
one can view the flow time as just another external scale, unrelated to the 
renormalization scale $\mu$. In this case, it is natural to modify the 
renormalization group equations to account for the change in the Green 
functions as we change both $\mu$ and $\tau$:
\begin{equation}
\mu\frac{\mathrm{d}}{\mathrm{d} \mu}\rightarrow  
\mu\frac{\mathrm{d}}{\mathrm{d} 
\mu}- 2\tau \frac{\mathrm{d}}{\mathrm{d} \tau}.
\end{equation}
Our choice of differential operator is constrained by the mass dimension of 
each scale, $\mu$ and $\tau$, but is not unique. In particular, one could 
choose 
the operator $\mu\mathrm{d}/\mathrm{d} 
\mu + \tau \mathrm{d}/\mathrm{d} \tau$. This freedom does not affect 
the logic of our discussion nor our conclusion, because alternative conventions 
can be absorbed into the definition of the renormalization parameters and 
anomalous dimensions in, for example, Eq.~(58).

At this stage it is worth commenting on the two scales in the problem, $\mu$ 
and $\tau$. In DIS, the spacetime separation of the corresponding
OPE, $x$, and renormalization scale $\mu$ are, in principle, two distinct 
scales. The spacetime separation is provided by the inverse momentum transfer 
of a particular DIS experiment or set of experiments. The renormalization 
scale, however, is a theoretical choice, and ultimately physical quantities 
should not depend on the renormalization scale. It is generally convenient to 
choose $\mu = 1/x$, but it is not strictly necessary.

The relationship between the flow time and the renormalization scale is 
analogous and these two scales are distinct. For the sOPE, the flow time can be 
considered as simply an external scale, imposed by some particular lattice 
``experiment'', and the renormalization scale is a convenient theoretical 
choice. We will see that it is helpful to tie these scales 
together, to reduce the two-scale problem to a single scale, but this is not 
formally necessary. Therefore, in the following analysis, the flow and 
renormalization scale should be understood as completely independent scales.

Considering again the sOPE for the two-point function, Eq.~
\eqref{eq:sope2pt},
we determine the renormalization group equation for the smeared Wilson 
coefficient, 
$d_{\rho^2}$, by following a procedure analogous to that outlined above. 

We assume that we are working in the small flow-time regime, which allows us 
to relate 
operators at vanishing and nonvanishing flow time 
\cite{Makino:2014taa,Suzuki:2013gza,Luscher:2011bx}:
\begin{equation}\label{eq:smallt}
[\phi^2(0)]_{\mathrm{R}}  =   {\cal 
Z}_{\rho^2}(\tau,\mu)\rho^2(\tau,0) + {\cal O}(\tau).
\end{equation}
This coefficient satisfies
\begin{equation}\label{eq:zgamma}
\mu\frac{\mathrm{d}}{\mathrm{d}\mu}\log({\cal 
Z}_{\rho^2}(\tau,\mu^2)) = 2\gamma_m,
\end{equation}
and to one loop is given by
\begin{equation}
{\cal Z}_{\rho^2}(\tau,\mu^2)= 
1-\frac{\lambda}{32\pi^2}\left[1+\gamma_{\;\mathrm{E}}+\log(2\tau 
\mu^2)\right]+{\cal 
O}(\lambda^2,\tau).
\end{equation}

We apply the renormalization group operator
\begin{equation}
\mu\frac{\mathrm{d}}{\mathrm{d} \mu} 
-2\tau\frac{\mathrm{d}}{\mathrm{d} \tau} + \left(N+2\right)\gamma
\end{equation}
to matrix elements of the operators in Eq.~\eqref{eq:smallt} coupled to 
$N$ 
external scalar fields to obtain
\begin{equation}\label{eq:srge}
\left[\mu\frac{\mathrm{d}}{\mathrm{d} 
\mu} -2\tau\frac{\mathrm{d}}{\mathrm{d} 
\tau} +2(\zeta_{\rho^2}-\gamma)\right]d_{\rho^2} = {\cal O}(\tau).
\end{equation}
Here
\begin{equation}
\zeta_{\rho^2} = \tau\frac{\mathrm{d}}{\mathrm{d}\tau}\log({\cal 
Z}_{\rho^2}(\tau,\mu^2))\label{eq:zetadef}
\end{equation}
is an anomalous dimension associated with the flow-time dependence of the 
operator $\rho^2(\tau,0)$. The renormalization group equation for the 
corresponding
matrix element of $\rho^2(\tau,0)$ coupled to $N$ external fields is 
given by
\begin{align}\label{eq:rhotau}
\bigg[\mu\frac{\mathrm{d}}{\mathrm{d} 
\mu}{} & -2\tau\frac{\mathrm{d}}{\mathrm{d} 
\tau} +2\zeta_{\rho^2}+N\gamma\bigg] \nonumber\\
{} & \times\big\langle \Omega 
|\rho^2(\tau,0)\widetilde{\phi}(p_1)\ldots\widetilde{\phi}(p_N) 
| \Omega \big\rangle = {\cal O}(\tau).
\end{align}
We note that this equation only holds provided the flow time is small compared 
with the momenta of the external particles, which for DIS would be of the order 
of hadronic scales.

If we now demand that the smearing scale, 
$\tau$, and the inverse of the renormalization scale, $\mu$, are 
proportional to each other, \emph{i.e.}, $\tau = b/\mu^2$ with $b$ real, then 
the 
renormalization group 
equation 
becomes
\begin{align}\label{eq:rhotau_mutau}
{} &
\bigg[2\mu\frac{\mathrm{d}}{\mathrm{d} 
\mu} +2\zeta_{\rho^2}+N\gamma\bigg] \nonumber\\
{} &\qquad \times\big\langle \Omega 
|\rho^2\left( b/\mu^2,0\right)\widetilde{\phi}(p_1)\ldots\widetilde{\phi
} (p_N) 
| \Omega \big\rangle = {\cal O}(b).
\end{align}
This renormalization group equation provides the starting point for a 
nonperturbative 
step-scaling method \cite{Monahan:2013lwa,Luscher:1991oxn} that evolves 
nonperturbative matrix elements to a high scale, where they can be combined 
with 
perturbative smeared Wilson coefficients. Here the renormalization group 
equation holds in 
the small flow time limit, or in other words, provided $b\ll 1$. This 
constraint automatically ensures that the flow time is also smaller than any 
hadronic length scales, $\Lambda_{\mathrm{QCD}} \ll \mu^2 \ll 1/\tau$, and is 
generally true for practical step-scaling methods 
\cite{Monahan:2013lwa,Luscher:1991oxn}. We are currently 
investigating this approach in QCD.

\section{Conclusion}

We have proposed a new method, the smeared operator product expansion (sOPE), 
to extract matrix elements from numerical nonperturbative calculations without 
power divergent mixing. The smeared operator product expansion is a general 
framework relevant to any asymptotically free theory with nonperturbative 
matrix elements that suffer from power divergent mixing. Within QCD, the most 
obvious application is to deep inelastic scattering, but other applications 
include nonperturbative determinations of $K\rightarrow \pi\pi$ decays 
\cite{Dawson:1997ic,Detmold:2005gg} and $B$-meson mixing 
\cite{Monahan:2014xra}. Beyond QCD, applications include nonperturbative 
studies 
of critical phenomena in the Heisenberg model, spin systems and other condensed 
matter systems.

In the sOPE, we expand nonlocal operators in a basis of smeared operators, 
the matrix elements of which can be determined on the lattice. We implement the 
smearing via the gradient flow, a 
classical evolution of the theory in a new dimension that smooths ultraviolet 
fluctuations. The continuum limit of these matrix elements is free of power 
divergent mixing, provided the localization scale, the smearing length, is kept 
fixed in the continuum limit. The resulting matrix elements are functions of 
two 
scales, the renormalization scale and the smearing length. The sOPE 
systematically relates these matrix elements to smeared Wilson coefficients, 
which can be calculated in perturbation theory, thereby providing a complete 
determination of the nonlocal operators.

% Use the proper section head for acknowledgments.
\begin{acknowledgments}
The authors would like to thank Martin L\"uscher for helpful discussions during 
the course of this work and Andrea Shindler for discussions regarding related 
work. This project was supported in part by the
U.S.~Department of Energy, Grant No.~DE-FG02-04ER41302. K.O.~was also  
supported by the U.S. Department of Energy through Grant No.~DE-AC05-06OR23177, 
under which JSA operates the Thomas Jefferson National 
Accelerator Facility. C.J.M.~was supported in part by the 
U.S.~National Science Foundation under Grant No.~NSF PHY10-034278.
\end{acknowledgments}

%

% End document
\end{document}